\documentclass[utf8]{FrontiersinHarvard}
\usepackage{physics}
\usepackage{amsmath,amsfonts,amssymb}
\usepackage{graphicx}
\usepackage{hyperref}
\usepackage{siunitx} 
\usepackage{url,lineno,microtype,subcaption}
\usepackage[onehalfspacing]{setspace}

\def\keyFont{\fontsize{8}{11}\helveticabold }
\def\Authors{Ran Guo}
\def\firstAuthorLast{}
\def\Address{Department of Physics, College of Science, Civil Aviation University of China, Tianjin 300300, China}
\def\corrAuthor{Ran Guo}
\def\corrEmail{rguo@cauc.edu.cn}

\begin{document}
\onecolumn

\title[Suprathermal effects on electron-acoustic instabilities]{Effects of suprathermal electrons on electron-acoustic instabilities} 
\author{\Authors}
\address{}
\correspondance{}
\extraAuth{}

\maketitle

\begin{abstract}
    We study the electron-acoustic instabilities in plasmas with two kappa-distributed electrons and stationary ions.
    The instabilities are driven by the relative drift between two electron components.
    The suprathermal effects of different species on growth rates and instability thresholds are analyzed and compared by numerical calculations.
    The present study reveals that the suprathermal electrons with slow most probable speed play more important roles than the suprathermal ones with fast most probable speed.
    The former significantly reduces the instabilities and raises the instability thresholds.
    The parameters used in this study are inspired from the observations in Earth's magnetopause.
\tiny
 \keyFont{ \section{Keywords:} plasma instabilities, suprathermal particles, electron-acoustic waves, particle velocity distribution, space plasmas}
\end{abstract}

\section{Introduction}
\label{sec:intro}
The suprathermal electrons are ubiquitous in non-equilibrium space plasmas.
Many observations indicate that these non-thermal electrons could be modeled very well by the kappa distribution in various plasma systems, such as solar winds \citep{Pierrard2016,Lazar2017a}, discrete auroral arcs \citep{Ogasawara2017},  the planetary magnetosphere \citep{Hapgood2011,Dialynas2009}, and cometary plasmas \citep{Broiles2016,Myllys2019}. 
The three-dimensional kappa velocity distribution is usually written as \citep{Pierrard2016,Lazar2017a},
\begin{equation}
    f_\kappa(\vb{v})=\frac{1}{(\kappa\pi\theta^2)^{3/2}} \frac{\Gamma(\kappa+1)}{\Gamma(\kappa-\frac{1}{2})}\left( 1 + \frac{v^2}{\kappa \theta^2} \right)^{-\kappa-1},
    \label{eq:kappa-pdf}
\end{equation}
where $\theta$ is the most probable speed (MPS) related to the kinetic temperature by,
\begin{equation}
    \theta = \sqrt{\frac{\kappa-\frac{3}{2}}{\kappa} \frac{2k_BT}{m}},
    \label{eq:theta}
\end{equation}
according to the definition $k_B T = \int \frac{1}{3} mv^2 f_\kappa \dd{\vb{v}}$.
The kappa distribution \eqref{eq:kappa-pdf} converges to the Maxwellian one when the parameter $\kappa$ goes to infinity.
Thus, a finite kappa index measures the distance divergent from the Maxwellian equilibrium.
Although the kappa distribution has been widely used in the literature, the formation mechanism is still under discussion \citep{Yoon2014,Livadiotis2019b,Guo2020,Guo2021}.

The suprathermal particles play an important role in affecting the physical properties of plasmas, such as the Debye length \citep{Hatami2018,Livadiotis2019}, the transport coefficients \citep{Wang2018,Guo2019,Husidic2021}, and collision frequencies \citep{Wang2021}.
A number of waves in kappa-distributed plasmas behave very differently from those in thermal equilibrium. 
The dust ion-acoustic waves were studied in unmagnetized plasmas with kappa-distributed particles \citep{Baluku2015,Lazar2018}.
These works found that the dispersion and Landau damping of dust ion-acoustic waves are critically changed in the presence of suprathermal electrons and ions.
\citet{Arshad2015} investigated the Landau damping of Langmuir twisted waves in kappa-distributed plasmas.
They found that the damping of Langmuir waves in the planar electric field is stronger than those in the twisted electric field.

Among these waves, the electron-acoustic wave (EAW) and its corresponding instability explain several phenomena in space plasmas, such as the formation of broadband electrostatic noises \citep{Pottelette1999,Singh2001a} and electron heating in solar flaring loops \citep{Chen2020}. 
\citet{Gary1985} first studied the behaviors of EAW and its instabilities \citep{Gary1987} in Maxwellian plasmas with hot and cool electrons. 
\citet{Mace1999} investigated the suprathermal effects on EAWs by assuming the kappa-distributed hot electrons, 
which found that the suprathermalization would reduce the Landau damping. 
\citet{Danehkar2011} studied the electron-acoustic solitary waves with suprathermal hot electrons and found that the kappa index affected the shape of solitons. 
\citet{Baluku2011} developed these works by modeling both hot and cool electrons with kappa distributions. 
Their investigation showed that the hot-to-cool temperature ratio and hot electron density are the two main factors determining the damping rate. 
However, in contrast to the above studies, our recent work \citep{Guo2021a} implied that the decisive factor of the damping rate is not the temperature difference but the sufficient gap of MPSs between two kappa-distributed electrons. 
It showed that the weakly damped EAWs could propagate in plasmas with two kappa-distributed electrons having the same temperature but different kappa indices.

In this work, we study the suprathermal effects on the electron-acoustic instabilities (EAIs) driven by the relative drift between two kappa electrons.
The most interesting findings are the suprathermal effects for different species appear very distinct.
The paper is organized as follows. 
In section \ref{sec:model}, we introduce the theoretical model of EAW propagating in plasmas with two kappa-distributed electrons.
In section \ref{sec:sup_eff}, we analyze the suprathermal effects of each electron component and study the thresholds of relative drifts to excite unstable EAWs.
At last, we make conclusions and discussions in section \ref{sec:sum}.

\section{Model}
\label{sec:model}
We consider a collisionless and electrostatic plasma consisting of two electron components and one ion component.
The ions are assumed to be spatially uniform and static to provide a neutral charge background.
The two electron species are modeled by the three-dimensional kappa distributions \citep{Summers1991,Mace1995},
\begin{equation}
    f(\vb{v}) = \sum_{\sigma=s,f} \frac{n_\sigma}{(\kappa_\sigma\pi\theta_\sigma^2)^{3/2}} \frac{\Gamma(\kappa_\sigma+1)}{\Gamma(\kappa_\sigma-\frac{1}{2})} \left[ 1+ \frac{(\vb{v}-\vb{u}_\sigma)^2}{\kappa_\sigma \theta_\sigma^2} \right]^{-\kappa_\sigma-1},
    \label{eq:mul-kappa-dist}
\end{equation}
where, for $\sigma$ species, the number density is denoted by $n_\sigma$, the kappa index by $\kappa_\sigma$, the MPS by $\theta_\sigma$, and the drift speed by $\vb{u}_\sigma$. 
The kappa index $\kappa_\sigma$ has to be larger than $3/2$ to maintain the convergence of the second moment of the kappa distribution \eqref{eq:mul-kappa-dist} \citep{Livadiotis2010a}. 

The subscripts $\sigma = s,f$ distinguish between two different electron components, which we call slow- and fast-MPS electrons, respectively, with the assumption $\theta_s<\theta_f$.
These terminologies may be different from the other works studying EAWs.
In the literature, these two electrons are called cool and hot electrons due to their different temperatures.
However, our previous work \citep{Guo2021a} showed that the weakly damped EAWs could propagate even if the two electrons have the same temperatures in kappa-distributed plasmas.
The reason is that, in terms of Eq. \eqref{eq:theta}, the MPSs of different species could be different if the temperatures are the same and the kappa indices are not.
Moreover, the difference between the MPSs is the main factor determining the EAW damping rate rather than the temperature difference in kappa-distributed plasmas \citep{Guo2021a}.
Hence, it is not appropriate to differentiate between the two electrons by their temperatures.
The two populations are renamed slow- and fast-MPS electrons due to their different MPSs.
We must stress that the words "slow" and "fast" do not denote the bulk speed of electrons in the present paper.

The kinetic temperature can be derived,
\begin{equation}
    k_BT_\sigma = \frac{\kappa_\sigma}{\kappa_\sigma-\frac{3}{2}}\frac{m\theta_\sigma^2}{2}.
    \label{eq:mul-T}
\end{equation}
It is worth noting that the parameters $\kappa_\sigma$, $\theta_\sigma$, and $T_\sigma$ are related by Eq. \eqref{eq:mul-T}, resulting in that only two of them are independent variables.
The kappa index must be independent of other parameters, so either $T_\sigma$ or $\theta_\sigma$ depends on $\kappa$.
These two choices could be both valid but for different physical processes \citep{Hellberg2009,Yoon2014,Livadiotis2015a,Lazar2016}. 
\citet{Lazar2016} suggested that $\theta_\sigma$ should be a $\kappa$-independent parameter if the kappa distribution is formed due to some particle acceleration processes. 
Their work reveals that such a choice is suitable for studying the effects of suprathermal particles.
Hence, we treat the MPS $\theta_\sigma$ as a $\kappa$-independent parameter.
The suprathermalization in this work means that the electron distribution changes from the Maxwellian distribution to the kappa one.
In this process, the kappa index reduces from infinity (Maxwellian) to a finite value.
Meanwhile, we assume that the MPS $\theta_\sigma$ does not change.

As we know, the EAI would be triggered due to a non-zero relative drift between the two electron components.
\citet{Gary1987} pointed out two different types of drifts describing two different systems. 
The first case is that both the slow- and fast-MPS electrons have non-zero drift speeds relative to the stationary ions.
In this case, another electron-ion acoustic instability, besides EAI, would be excited because of the relative motion between drifting slow-MPS electrons and static ions. 
The second case is that the fast-MPS electrons drift with respect to the slow-MPS electrons and ions, but there is no drift between slow-MPS electrons and ions. 
In this case, only EAI would be triggered, leading to an unstable wave parallel to the drifting velocity of fast-MPS electrons.
The latter system is studied in this work to focus on the suprathermal effects on EAIs.
Furthermore, we choose the slow-MPS electrons as the reference frame, so the ions are immobile, the slow-MPS electrons has no drift ($\vb{u}_s=0$), and the fast-MPS electrons drift with a non-zero speed ($\vb{u}_f \ne 0$).
We emphasize that such a plasma system has a non-zero current carried by the drifting species. 

To analyze the suprathermal effects of slow- and fast-MPS electrons, we compare the EAIs in the following three models.
The first model is a contrast model in which the plasma consists of Maxwellian cool and hot electrons that are called the slow- and fast-MPS ones in the present study.
We denote it as the Maxwellian-Maxwellian (MM) model hereafter.
In MM plasmas, the two electrons can be differentiated by either the MPSs or the temperatures because the MPS \eqref{eq:theta} is directly determined by the temperature in the limit of $\kappa \rightarrow \infty$.
So, the slow- and fast-MPS electrons are just alternative names for the cool and hot ones in this case.
The second Maxwellian-kappa (MK) model is constituted by slow-MPS Maxwellian electrons and fast-MPS kappa-distributed ones.
Similarly, the third model, suprathermalizing only slow-MPS electrons, is denoted as the kappa-Maxwellian (KM) model.
As we proved in the previous studies \citep{Guo2021a}, the gap between the MPSs is the definitive factor for the EAW damping. 
One could infer a similar conclusion for the growth rate in EAI studies.
Therefore, we have to compare the three models with the same $\theta_s$ and $\theta_f$.
Under this condition, the suprathermalization would increase the temperature because a smaller kappa index results in a higher temperature in Eq. \eqref{eq:mul-T} if $\theta_\sigma$ is fixed.
We have to stress that the temperatures for two electrons are unnecessary to be the same in this work.

Some observations and simulations supported the existence of the plasmas studied in the present work, i.e., the system consisting of drifting fast-MPS electrons, non-drifting slow-MPS electrons, and immobile ions.
\citet{Ergun2016} studied the electrostatic waves observed at the Earth’s magnetopause, including the ion-acoustic, the electron-acoustic, and the beam mode. 
These waves were successfully explained by their simulations of the fast-MPS (hot) magnetosheath plasmas flowing into the slow-MPS (cool) magnetosphere ones.

The KM model is rare in the literature, but the plasma system described by such a model probably exists in space.
The suprathermal electrons were observed in the Earth's magnetosheath \citep{Kasaba2000} and magnetosphere \citep{Hapgood2011,Eyelade2021}, which could be modeled by the kappa distribution.
However, it does not mean that the electrons in the magnetosphere and magnetosheath can only be kappa-distributed;
those electrons can also be Maxwellian.
Thereby, it is entirely possible that the magnetosphere electrons are kappa-distributed and the magnetosheath electrons are Maxwellian.
In the magnetopause, the fast-MPS (hot) Maxwellian electrons in the magnetosheath might flow into the slow-MPS (cool) kappa-distributed electrons in the magnetosphere, which is just the system described by the KM model.

\section{Numerical Analysis}
\label{sec:sup_eff}
The linear dispersion relation of EAIs could be inferred from those of stable EAWs \citep{Guo2021a,Mace1995},
\begin{equation}
     1 + \sum_{\sigma=s,f} \frac{2 \omega_\sigma^2}{k^2 \theta_\sigma^2} \left[ 1 - \frac{1}{2\kappa_\sigma} +\xi_\sigma Z(\kappa_\sigma;\xi_\sigma) \right]=0,
     \label{eq:pdr}
\end{equation}
with the replacement $\xi_\sigma = (\omega-\vb{k}\cdot\vb{u}_\sigma)/(k \theta_\sigma)$. 
In the above equation, for $\sigma$ component, $\omega_\sigma = \sqrt{n_\sigma e^2/(m \varepsilon_0)}$ is the plasma frequency, $\vb{k}$ is the wave vector, and $Z(\kappa_\sigma;\xi_\sigma)$ is the modified plasma dispersion function given by \citep{Mace1995}, 
\begin{equation}
    Z(\kappa_\sigma;\xi_\sigma) = \frac{\Gamma(\kappa_\sigma)}{\sqrt{\pi\kappa_\sigma}\Gamma\left(\kappa_\sigma-\frac{1}{2}\right)} 
    \int_{-\infty}^{+\infty} \frac{\left( 1+\frac{s^2}{\kappa_\sigma} \right)^{-\kappa_\sigma-1}}{s-\xi_\sigma} \dd{s},
    \label{eq:Z-int}
\end{equation}
which can be re-written in the form of the hypergeometric function for numerical purposes \citep{Mace1995}, 
\begin{equation}
    Z(\kappa_\sigma;\xi_\sigma) = i \frac{\left( \kappa_\sigma + \frac{1}{2} \right)\left( \kappa_\sigma - \frac{1}{2} \right)}{\kappa_\sigma^{3/2}(\kappa_\sigma+1)} 
    {_2 F_1} \left[1,2\kappa_\sigma+2;\kappa_\sigma+2;\frac{1}{2}\left(1-\frac{\xi_\sigma}{i\sqrt{\kappa_\sigma}}  \right)\right].
    \label{eq:Z-2f1}
\end{equation}

The analytical growth rate is not derived here.
\citet{Baluku2011} indicated that the analytical damping rate of EAWs in kappa-distributed plasmas is highly complex, so it has little practical value.
We could infer that the growth rate of EAIs also has a complicated expression which provides limited information.

Thereby, the dispersion relation \eqref{eq:pdr} is solved numerically throughout this paper.
The hypergeometric function is calculated by mpmath, a Python library for floating-point arithmetic with arbitrary precision.
The codes of this study could be found at: \url{https://github.com/rguo1988/EAI-in-kappa-plasmas}.


\subsection{Wave frequency and Growth Rate}
The unstable wave propagates in the direction parallel to the drift speed $\vb{u}_f$, so we can only consider this parallel direction, reducing the system to a one-dimensional plasma.
\begin{figure}
	\centering
    \includegraphics[width=0.95\textwidth]{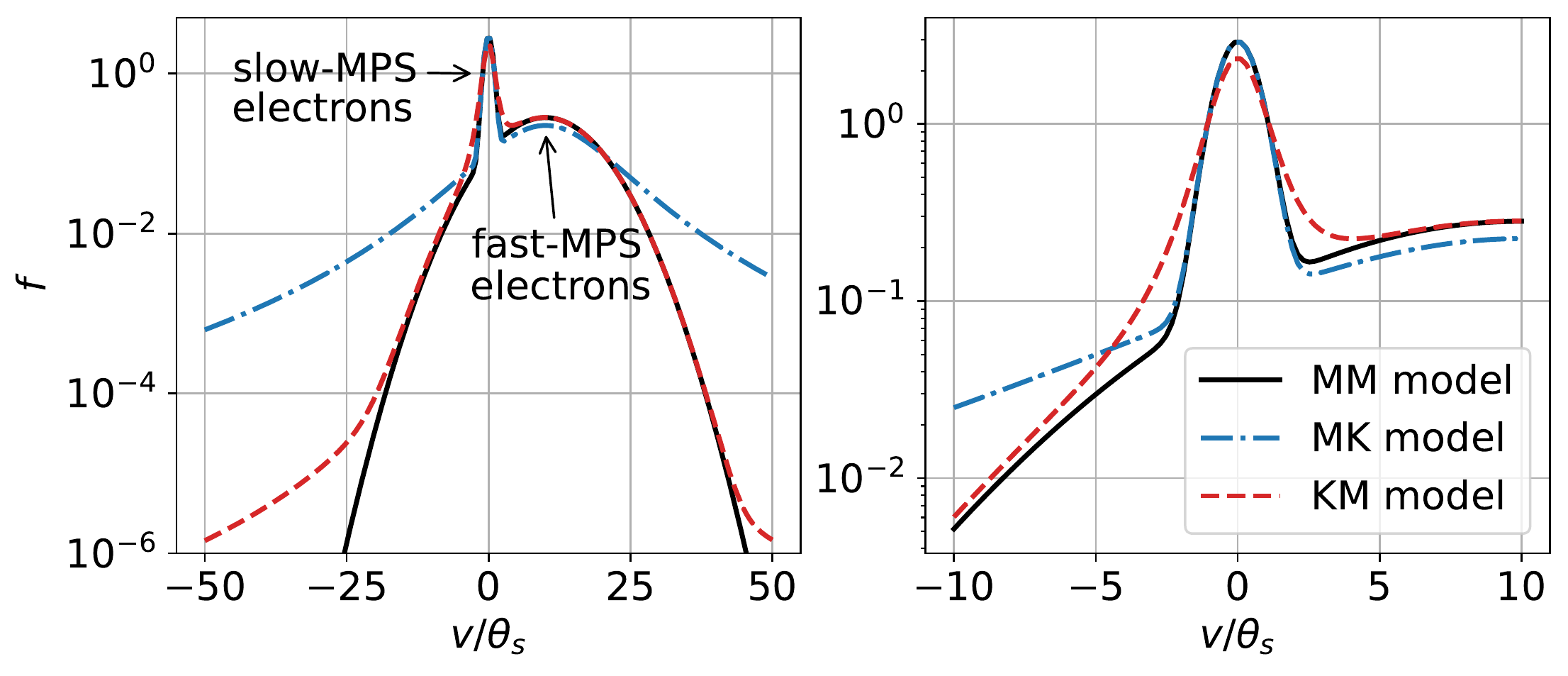}
    \caption{The one-dimensional velocity distribution for MM, MK, and KM models.
    The parameters $\kappa_s=\infty$ and $\kappa_f=2$ are set in MK model, while $\kappa_s=2$ and $\kappa_f=\infty$ in KM model.  
    For all three models, the MPSs are set as $\theta_s=0.1$ and $\theta_f=1$.
    The drift of fast-MPS species are $u_f/\theta_s=10$.
    The number densities are $n_s/n_0=n_f/n_0=0.5$, where $n_0 = n_s+n_f$ is the total electron density.
        }
    \label{fig:pdf}
\end{figure}
By integrating Eq. \eqref{eq:mul-kappa-dist}, the one-dimensional kappa distribution is,
\begin{equation}
    f_{\mathrm{1D}}(v) = \sum_{\sigma=s,f} \frac{n_\sigma}{\sqrt{\kappa_\sigma\pi\theta_\sigma^2}} \frac{\Gamma(\kappa_\sigma)}{\Gamma(\kappa_\sigma-\frac{1}{2})} \left[ 1+ \frac{(v-u_\sigma)^2}{\kappa_\sigma \theta_\sigma^2} \right]^{-\kappa_\sigma},
    \label{eq:one-kappa-dist}
\end{equation}
which is plotted in Fig. \ref{fig:pdf} for all three models.
The distributions are illustrated in a global velocity range on the left panel while magnified in a low velocity region on the right panel.
We set $\kappa_s=\infty$ and $\kappa_f=2$ for MK model, while $\kappa_s=2$ and $\kappa_f=\infty$ for KM model throughout this paper.  
In this figure, the MPSs are set as $\theta_s=0.1$ and $\theta_f=1$ for all three models.
The drift of fast-MPS species are $u_f/\theta_s=10$.
The number densities are $n_s/n_0=n_f/n_0=0.5$, where $n_0 = n_s+n_f$ is the total electron density.
The above parameters are inspired from the observations of space plasmas \citep{Eyelade2021, Ergun2016}.
In comparison with the MM model, the suprathermalization of fast-MPS electrons (MK model) changes the distribution in medium and high speed regions but does not change that in the low speed region, as shown by the blue dotted-dashed lines in Fig. \ref{fig:pdf}.
However, the suprathermal slow-MPS electrons play a different role.
The KM and MM distributions overlap with each other in the medium speed region but diverge in low and high speed regions, as shown by the red dashed lines in Fig. \ref{fig:pdf}.

\begin{figure}
	\centering
    \includegraphics[width=0.585\textwidth]{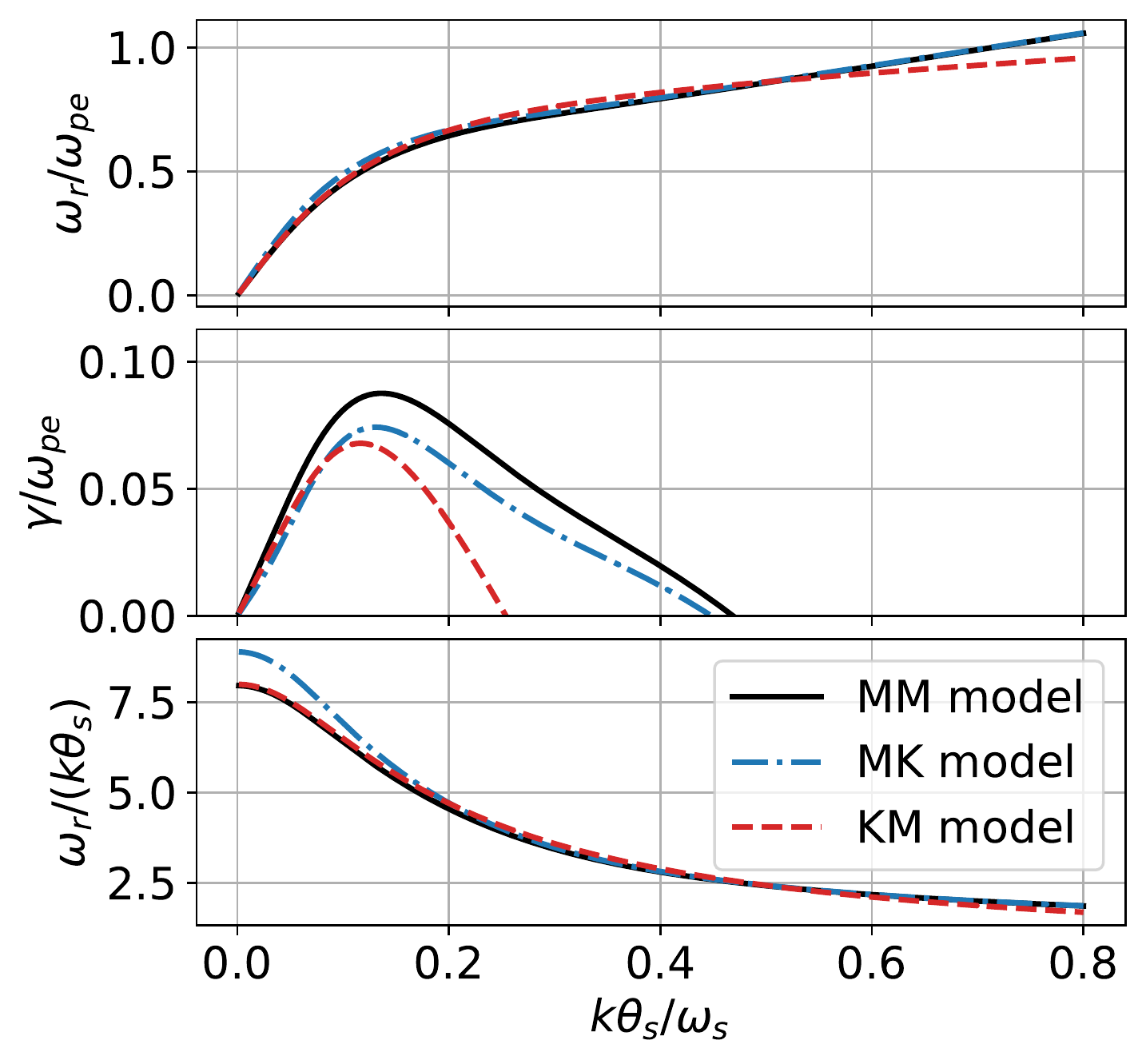}
    \caption{The wave frequency (upper panel), growth rate (middle panel), and wave speed (lower panel) of EAIs in the three models.
    The parameters are set as the same as those in Fig. \ref{fig:pdf}. 
    The real frequency $\omega_r$ and growth rate $\gamma$ are both scaled by the total plasma frequency $\omega_{pe} = \sqrt{\omega_s^2+\omega_f^2}$.
        }
    \label{fig:pdr}
\end{figure}

The real frequency, growth rate, and wave speed of EAIs are illustrated in Fig. \ref{fig:pdr}.
In this figure, the parameters of the three models are set as the same as those in Fig. \ref{fig:pdf}. 
The normalized wavenumber $k\theta_s/\omega_s$ is used rather than $k\lambda_{\kappa\sigma}$ where $\lambda_{\kappa\sigma}$ is the Debye length in kappa-distributed plasmas given by \citep{Livadiotis2019}, 
\begin{equation}
    \lambda_{\kappa \sigma} = \sqrt{\frac{\kappa_\sigma-\frac{3}{2}}{\kappa_\sigma-\frac{1}{2}}} \sqrt{\frac{\varepsilon_0 k_B T_\sigma}{n_\sigma e^2}}
    = \sqrt{\frac{\kappa_\sigma}{2\kappa_\sigma-1}}\frac{\theta_\sigma}{\omega_\sigma}.
    \label{eq:lambda}
\end{equation}
The reason is that the Debye length \eqref{eq:lambda} is $\kappa$-dependent, leading to different $\lambda_{\kappa\sigma}$ in three models.
One does not expect the dimensionless wavenumber changes due to the suprathermalizations of different species in comparisons.
Therefore, we use $k\theta_s/\omega_s$ as the normalized wavenumber, unchanged for varied kappa indices.

Figure \ref{fig:pdr} implies that the suprathermal effects of each species are distinct.
In the upper panel, we find the suprathermal slow-MPS electrons (KM model) affect the wave frequency to a small extent at long wavelengths but to a large extent at short wavelengths, as shown by the red dashed line.
However, the suprathermalization of fast-MPS electrons (MK model) only slightly alters the wave frequency in the long-wave region but has almost no effects in the short-wave area, as shown by the blue dotted-dashed line.
These suprathermal effects on the wave frequency are very similar to those of stable EAWs without drifting components, 
which has been already studied in previous works \citep{Baluku2011, Guo2021a}. 

In the middle panel of Fig. \ref{fig:pdr}, 
the EAIs would be declined by the suprathermalization of both slow- and fast-MPS electrons.
The suprathermal slow-MPS electrons (KM model) suppress the instabilities more than the fast-MPS ones (MK model) in large wavenumbers, but the conclusion is quite the contrary in small wavenumbers. 
This effect can be attributed to the derivatives of distributions at the wave speed.
As we know, the growth rate of an unstable wave has a positive correlation with the derivative of distributions at the wave speed.
From the lower panel of Fig. \ref{fig:pdr}, one finds the range of unstable EAW speed is roughly $2.5 \le \omega_r/(k\theta_s) \le 9$.
In such a velocity region, the suprathermalization of slow-MPS electrons would flatten the total distribution and lesson the velocity range with a positive slope, as shown by the red dashed line in the right panel of Fig. \ref{fig:pdf}.
On the contrary, one finds that the suprathermalization of fast-MPS electrons modifies the slope of the total distribution moderately and hardly alters the positive slope region.
\begin{figure}
	\centering
    \includegraphics[width=0.585\textwidth]{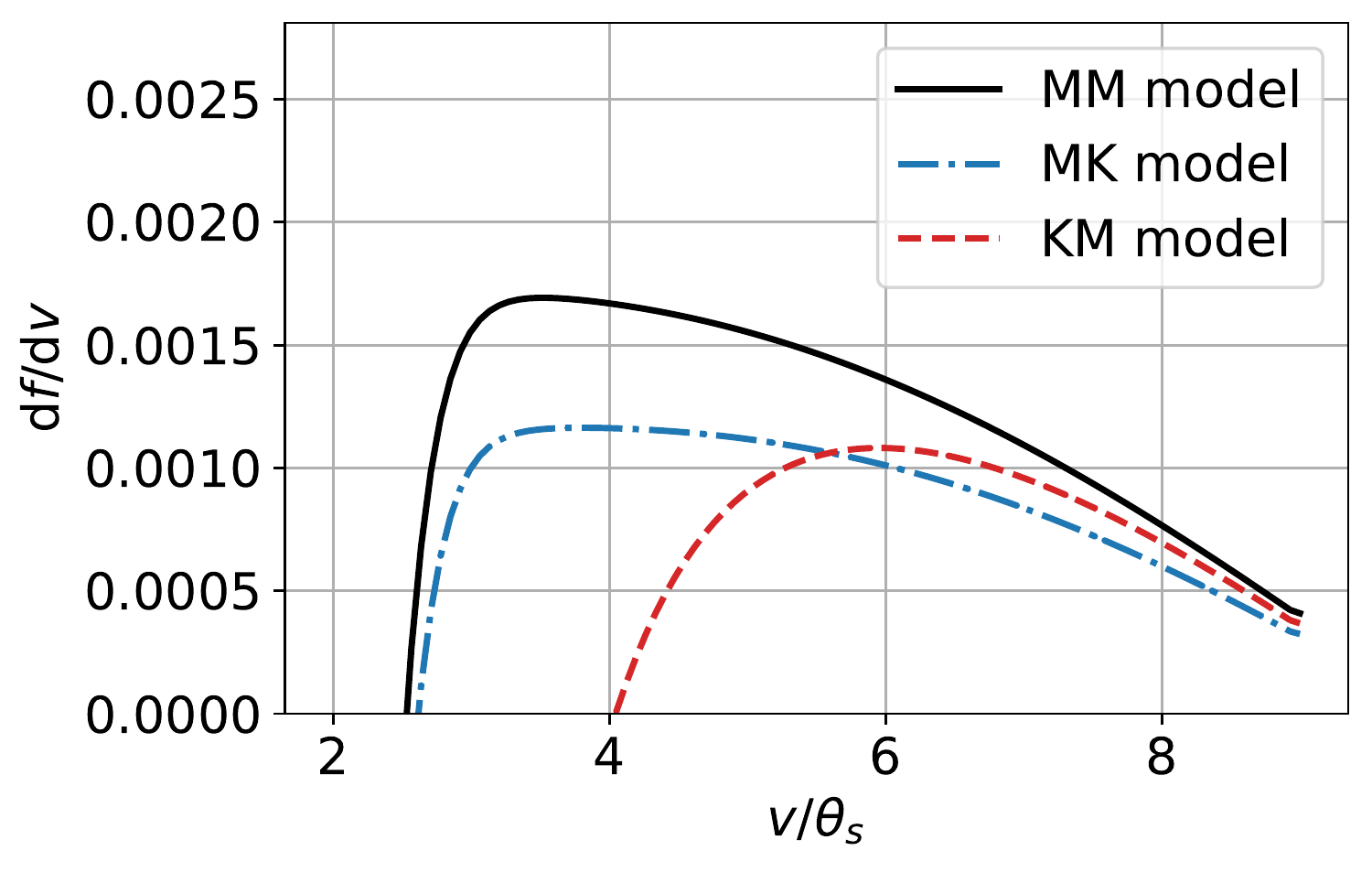}
    \caption{The derivatives of velocity distributions for three models in the range of unstable EAW speeds.
    The parameters are set as the same as those in Fig. \ref{fig:pdf}. 
        }
    \label{fig:dev-pdf}
\end{figure}
It could be presented more clearly by the derivatives of electron distributions \eqref{eq:one-kappa-dist},
\begin{equation}
    \dv[]{f_{\mathrm{1D}}}{v} = -\sum_{\sigma=s,f} \frac{n_\sigma}{\sqrt{\kappa_\sigma\pi\theta_\sigma^2}} \frac{\Gamma(\kappa_\sigma)}{\Gamma(\kappa_\sigma-\frac{1}{2})} \left[ 1+ \frac{(v-u_\sigma)^2}{\kappa_\sigma \theta_\sigma^2} \right]^{-\kappa_\sigma-1} \frac{2(v-u_\sigma)}{\theta_\sigma^2},
    \label{eq:one-kappa-dist-dev}
\end{equation}
which are drawn for all three models in Fig. \ref{fig:dev-pdf} with the same parameters used in Fig. \ref{fig:pdf}.
Figure \ref{fig:dev-pdf} illustrates that both the KM and MK models lead to decreased derivatives by comparison with the MM model.
The reduction of the positive derivatives implies that the EAWs in KM and MK models have a smaller growth rate than the MM model.
In addition, the KM model has a smaller derivative than the MK one for a slow EAW but a larger derivative for a fast EAW.
Consequently, the slow EAWs are more stable in the KM model than the MK model, while the fast EAWs have the opposite conclusion, as shown in the middle and lower panel of Fig. \ref{fig:pdr}. 

\subsection{EAI threshold}
\begin{figure}
	\centering
    \includegraphics[width=0.585\textwidth]{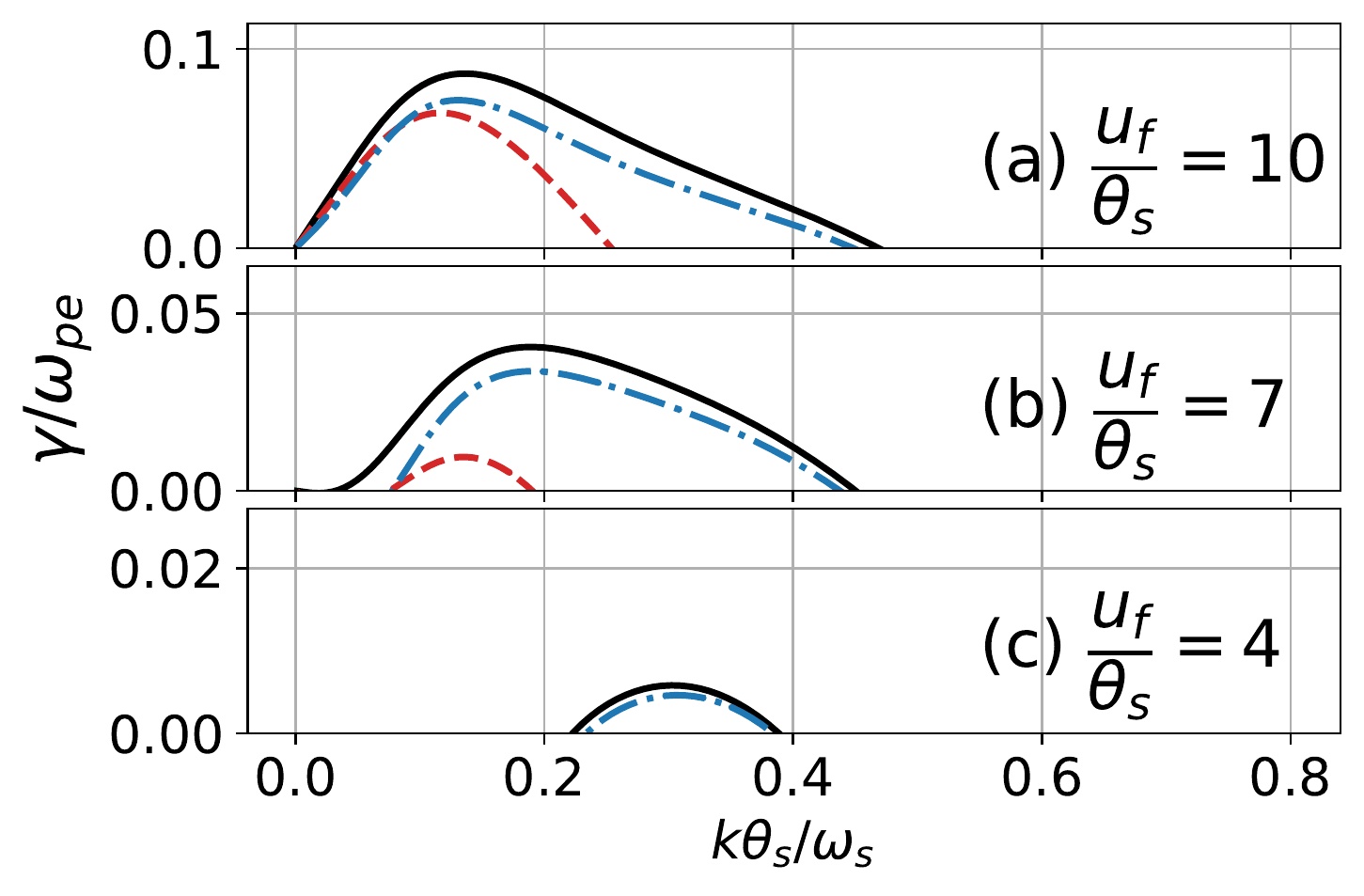}
    \caption{EAW growth rate for varied drift speeds.
    The kappa indice of three models, the MPSs, and the number densities are the same as those in Fig. \ref{fig:pdf}. 
    The legend is as same as Fig. \ref{fig:pdr}.
    }
    \label{fig:eai_a}
\end{figure}

The EAI is excited by the relative drift between two electron species.
We plot the EAW growth rate with different drift speeds $u_f/\theta_s=10,7$ and $4$ in Fig. \ref{fig:eai_a}.
The kappa indices of the three models, the MPSs, and the number densities are the same as those in Fig. \ref{fig:pdf}. 
It shows that the maximum $\gamma/\omega_{pe}$ decreases during the diminution of the drifting speed $u_f/\theta_s$.
The suprathermalization of slow-MPS electrons (red dashed line) accelerates this process dramatically, but the suprathermal fast-MPS electrons (blue dotted-dashed line) affect barely.
It suggests the suprathermal slow-MPS electron would raise the instability threshold significantly.
\begin{figure}
	\centering
    \includegraphics[width=0.585\textwidth]{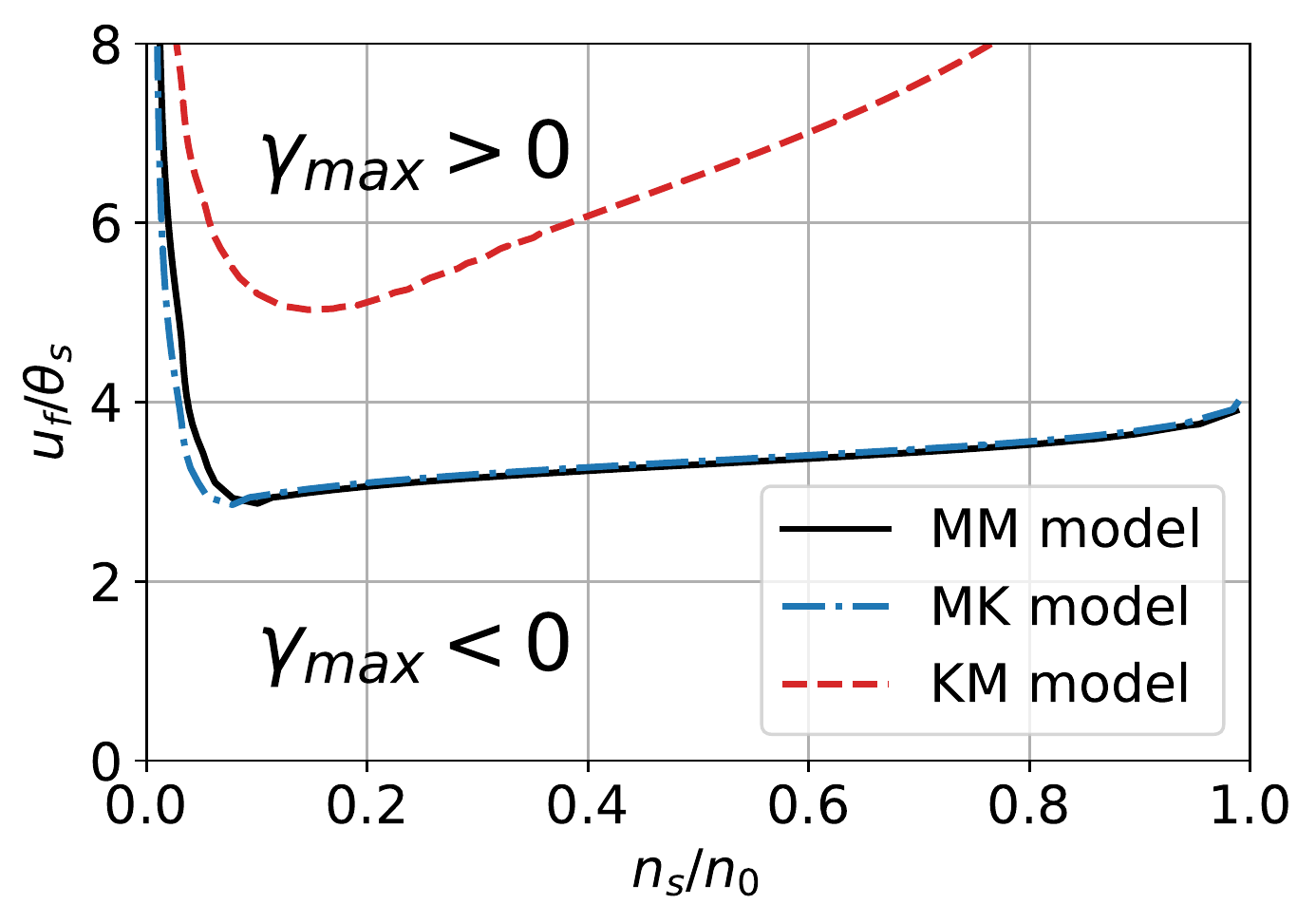}
    \caption{EAI threshold of the relative drift $u_f/\theta_s$ for different fractions of slow-MPS electrons $n_s/n_0$ in three models.
    The MPS ratio $\theta_f/\theta_s=10$ is fixed in this figure.
        }
    \label{fig:threshold-a-n}
\end{figure}
\begin{figure}
	\centering
    \includegraphics[width=0.585\textwidth]{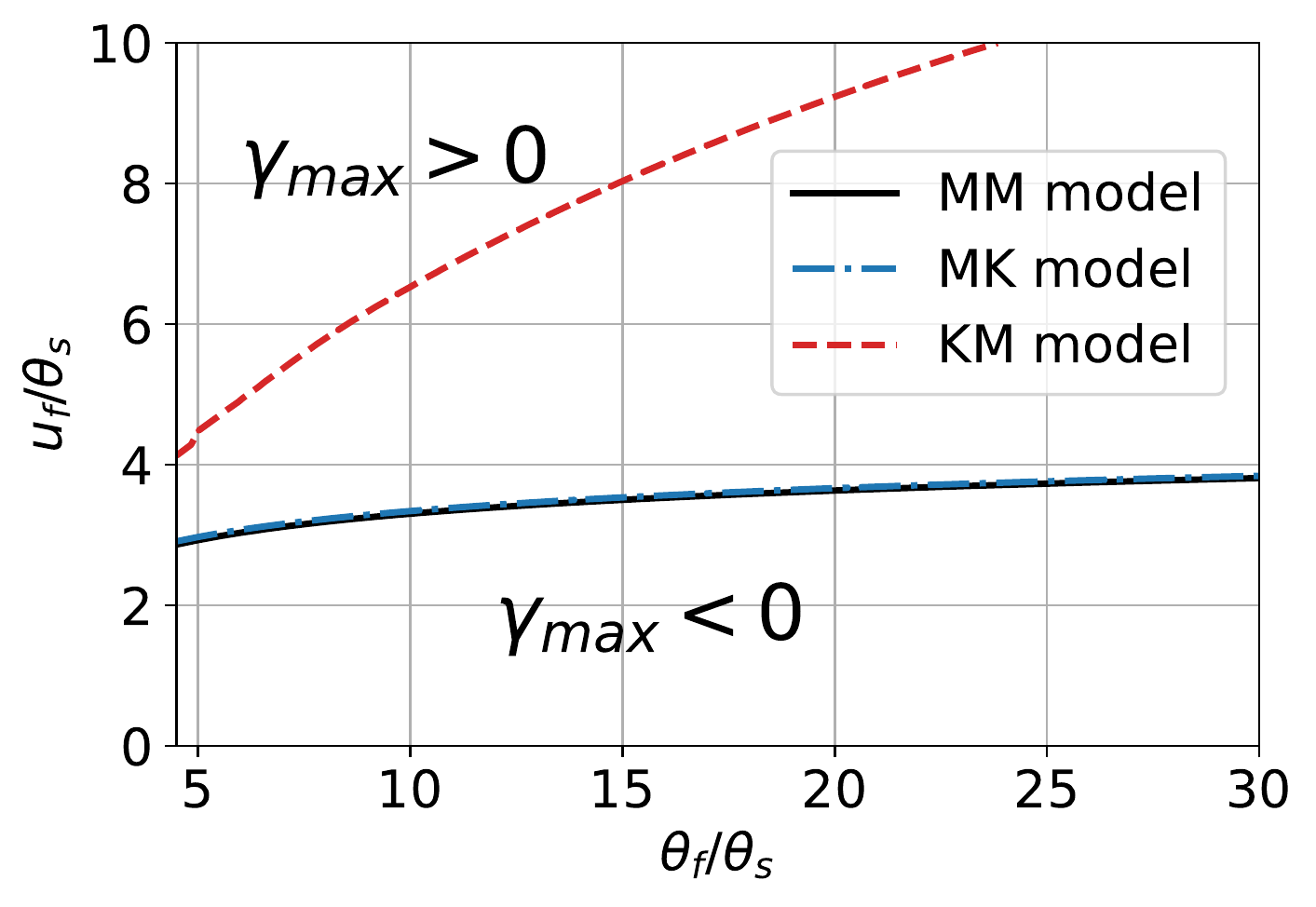}
    \caption{EAI threshold of the relative drift $u_f/\theta_s$ for different MPS ratios $\theta_f/\theta_s$ in three models.
    The fraction of slow-MPS electrons $n_s/n_0=0.5$ is fixed in this figure.
        }
    \label{fig:threshold-a-vfvs}
\end{figure}
Figures \ref{fig:threshold-a-n} and \ref{fig:threshold-a-vfvs} are plotted to illustrate the threshold of drift speeds for different slow-MPS component fractions $n_s/n_0$ and the MPS ratios $\theta_f/\theta_s$.
Figure \ref{fig:threshold-a-n} indicates that the suprathermalization of slow-MPS electrons (KM model) narrows the parameter space of the instabilities obviously, as we expected.
Furthermore, it is also not surprising that the EAI in the KM model is harder to trigger with an increasing fraction of slow-MPS electrons because it is the suprathermal slow-MPS electrons that enhance the EAI threshold.
However, in the case of the MK model, the suprathermalization slightly changes the unstable region.
Therefore, the suprathermal fast-MPS electrons only have a little influence.
Figure \ref{fig:threshold-a-vfvs} shows similar results that the instability separatrix is shifted upwards by suprathermalizing slow-MPS electrons (KM model) but is nearly unchanged by suprathermalizing fast-MPS ones (MK model).
The suprathermalization of slow-MPS electrons raises the instability threshold rapidly when the gap between the MPSs enlarges.
It could be explained as follows.
When $\theta_f/\theta_s$ increases, the EAW speed appears in the tail of the slow-MPS electron distribution, where more particles are suprathermalized and thus raise the instability threshold.

It is worth noting that most results from the earlier work \citep{Gary1987} were based on the zero-current system (Figs. 4-8 in Gary's paper).
It is different from the plasma system considered in the present study \citep[see][]{Gary1987,Gary1993}. 
Nevertheless, Fig. 2 in Gary's work \citep{Gary1987} was calculated in the same system as the present study.
In that figure, Gary showed that the approximate instability threshold is $v_0/v_c\approx 5$, where $v_0$ is the drifting speed of hot electrons and $v_c=\sqrt{T_c/m}$ is the thermal speed of cool species.
This threshold is calculated with the parameters $n_c/n_0=n_h/n_0=0.5$ and $T_h/T_c=100$, where $n_c$ ($n_h$) is the number density and $T_c$ ($T_h$) is the temperature for cool (hot) electrons.
In our notations, Gary's threshold is rewritten as $u_f/\theta_s \approx 3.5$ for $n_s/n_0=n_f/n_0=0.5$ and $\theta_f/\theta_s=10$, which is consistent with our results of the MM model in Fig. \ref{fig:threshold-a-n}.

The results of two suprathermal electrons, i.e., the kappa-kappa model, could be estimated qualitatively from those of the KM and MK models.
On the one hand, the dispersion relation \eqref{eq:pdr} implies that the suprathermal effects of both components could be regarded as a sum of the suprathermal effects of each species.
On the other hand, the suprathermal electrons with slow MPS significantly change the instability threshold, while the suprathermal fast-MPS ones affect the threshold only to a small extent.
Therefore, the results of the KK model must be very close to those of the KM model, just like the results of the MK and MM models shown in Figs. \ref{fig:threshold-a-n} and \ref{fig:threshold-a-vfvs}.

\section{Summary and Discussion}
\label{sec:sum}
In this work, we investigate EAIs in plasmas with two kappa-distributed electrons and static ions.
The two electrons are called the slow- and fast-MPS ones in terms of the different MPSs.
We assume the slow-MPS electrons and immobile ions have no drift, and the instabilities are triggered by the drifting fast-MPS electrons.
To analyze the suprathermal effects of different species, we study three models in which the two electrons follow MM, MK, and KM distributions, respectively.
By comparing the growth rates in three models, we find both the suprathermalizations of slow- and fast-MPS electrons reduce the instabilities.
The results are illustrated in Fig. \ref{fig:pdr}.
Further, the slow-MPS suprathermal electrons suppress the unstable wave more strongly than the fast-MPS ones in large wavenumbers but play the opposite role in small wavenumbers.
Besides, the suprathermalizations also affect the threshold of the drifting speed exciting the instabilities, as shown in Figs. \ref{fig:threshold-a-n} and \ref{fig:threshold-a-vfvs}.
It implies that the suprathermalizations of slow-MPS electrons significantly raise the instability threshold while the suprathermal fast-MPS electrons affect the threshold only to a small extent.

The present investigation may be used to study the behaviors of real space plasmas.
For instance, we consider the processes that the cool (slow-MPS) magnetosphere plasmas mix with the hot (fast-MPS) magnetosheath plasmas, which could occur in magnetopause \citep{Ergun2016}.
In terms of the observation data \citep{Ergun2016}, the typical parameters for such processes are $n_s \sim \SI{1}{cm^{-3}}$, $n_f \sim \SI{10}{cm^{-3}}$, $T_s \sim \SI{1}{eV}$, and $T_f \sim \SI{100}{eV}$.
This parameter set leads to $n_s/n_0 \sim 0.1$, and $\theta_f/\theta_s \sim 10$ for the MM model, $\theta_f/\theta_s \sim 5$ for the MK model, and $\theta_f/\theta_s \sim 20$ for the KM model.
Therefore, the parameters used in Sec. \ref{sec:sup_eff} are in the same order of magnitude as the actual parameters.
Although the magnetosheath and magnetosphere are magnetized plasmas, our work can be applied to analyze the electrostatic waves parallel to the magnetic field.

\section*{Funding}
This work was supported by the National Natural Science Foundation of China (No.12105361) and by the Supporting Fund from Civil Aviation University of China (No.3122022PT18).


\bibliographystyle{Frontiers-Harvard}
\bibliography{refs}

\begin{thebibliography}{39}
\providecommand{\natexlab}[1]{#1}
\expandafter\ifx\csname urlstyle\endcsname\relax
  \providecommand{\doi}[1]{doi:\discretionary{}{}{}#1}\else
  \providecommand{\doi}{doi:\discretionary{}{}{}\begingroup
  \urlstyle{rm}\Url}\fi
\providecommand{\selectlanguage}[1]{\relax}
\providecommand{\bibAnnoteFile}[1]{%
  \IfFileExists{#1}{\begin{quotation}\noindent\textsc{Key:} #1\\
  \textsc{Annotation:}\ \input{#1}\end{quotation}}{}}
\providecommand{\bibAnnote}[2]{%
  \begin{quotation}\noindent\textsc{Key:} #1\\
  \textsc{Annotation:}\ #2\end{quotation}}

\bibitem[{Arshad et~al.(2015)Arshad, ur~Rehman, and Mahmood}]{Arshad2015}
Arshad, K., ur~Rehman, A., and Mahmood, S. (2015).
\newblock Landau damping of langmuir twisted waves with kappa distributed
  electrons.
\newblock \emph{Phys Plasmas} 22, 112114.
\newblock \doi{10.1063/1.4935845}
\bibAnnoteFile{Arshad2015}

\bibitem[{Baluku and Hellberg(2015)}]{Baluku2015}
Baluku, T.~K. and Hellberg, M.~A. (2015).
\newblock Kinetic theory of dust ion acoustic waves in a kappa-distributed
  plasma.
\newblock \emph{Phys Plasmas} 22, 083701.
\newblock \doi{10.1063/1.4927581}
\bibAnnoteFile{Baluku2015}

\bibitem[{Baluku et~al.(2011)Baluku, Hellberg, and Mace}]{Baluku2011}
Baluku, T.~K., Hellberg, M.~A., and Mace, R.~L. (2011).
\newblock Electron acoustic waves in double-kappa plasmas: Application to
  saturn's magnetosphere.
\newblock \emph{J. Geophys. Res. Space Phys.} 116, A04227.
\newblock \doi{10.1029/2010ja016112}
\bibAnnoteFile{Baluku2011}

\bibitem[{Broiles et~al.(2016)Broiles, Burch, Chae, Clark, Cravens, Eriksson
  et~al.}]{Broiles2016}
Broiles, T.~W., Burch, J.~L., Chae, K., Clark, G., Cravens, T.~E., Eriksson,
  A., et~al. (2016).
\newblock Statistical analysis of suprathermal electron drivers at
  67p/churyumov{\textendash}gerasimenko.
\newblock \emph{Mon. Not. R. Astron. Soc.} 462, S312--S322.
\newblock \doi{10.1093/mnras/stw2942}
\bibAnnoteFile{Broiles2016}

\bibitem[{Chen et~al.(2020)Chen, Wu, Xiang, Shi, Ma, Tang et~al.}]{Chen2020}
Chen, L., Wu, D.~J., Xiang, L., Shi, C., Ma, B., Tang, J.~F., et~al. (2020).
\newblock The electron acoustic wave and its role in solar flaring loops
  heating.
\newblock \emph{Astrophys. J.} 904, 193.
\newblock \doi{10.3847/1538-4357/abc00b}
\bibAnnoteFile{Chen2020}

\bibitem[{Danehkar et~al.(2011)Danehkar, Saini, Hellberg, and
  Kourakis}]{Danehkar2011}
Danehkar, A., Saini, N.~S., Hellberg, M.~A., and Kourakis, I. (2011).
\newblock Electron-acoustic solitary waves in the presence of a suprathermal
  electron component.
\newblock \emph{Phys Plasmas} 18, 072902.
\newblock \doi{10.1063/1.3606365}
\bibAnnoteFile{Danehkar2011}

\bibitem[{Dialynas et~al.(2009)Dialynas, Krimigis, Mitchell, Hamilton, Krupp,
  and Brandt}]{Dialynas2009}
Dialynas, K., Krimigis, S.~M., Mitchell, D.~G., Hamilton, D.~C., Krupp, N., and
  Brandt, P.~C. (2009).
\newblock Energetic ion spectral characteristics in the saturnian magnetosphere
  using cassini/mimi measurements.
\newblock \emph{J. Geophys. Res. Space Phys.} 114, A01212.
\newblock \doi{10.1029/2008ja013761}
\bibAnnoteFile{Dialynas2009}

\bibitem[{Ergun et~al.(2016)Ergun, Holmes, Goodrich, Wilder, Stawarz, Eriksson
  et~al.}]{Ergun2016}
Ergun, R.~E., Holmes, J.~C., Goodrich, K.~A., Wilder, F.~D., Stawarz, J.~E.,
  Eriksson, S., et~al. (2016).
\newblock Magnetospheric multiscale observations of large-amplitude, parallel,
  electrostatic waves associated with magnetic reconnection at the
  magnetopause.
\newblock \emph{Geophys Res Lett} 43, 5626--5634.
\newblock \doi{10.1002/2016gl068992}
\bibAnnoteFile{Ergun2016}

\bibitem[{Eyelade et~al.(2021)Eyelade, Stepanova, Espinoza, and
  Moya}]{Eyelade2021}
Eyelade, A.~V., Stepanova, M., Espinoza, C.~M., and Moya, P.~S. (2021).
\newblock On the relation between kappa distribution functions and the plasma
  beta parameter in the earth's magnetosphere: {THEMIS} observations.
\newblock \emph{Astrophys. J. Suppl. Ser.} 253, 34.
\newblock \doi{10.3847/1538-4365/abdec9}
\bibAnnoteFile{Eyelade2021}

\bibitem[{Gary(1987)}]{Gary1987}
Gary, S.~P. (1987).
\newblock The electron/electron acoustic instability.
\newblock \emph{Phys Fluids} 30, 2745--2749.
\newblock \doi{10.1063/1.866040}
\bibAnnoteFile{Gary1987}

\bibitem[{Gary(1993)}]{Gary1993}
Gary, S.~P. (1993).
\newblock \emph{Theory of Space Plasma Microinstabilities} (Cambridge
  University Press).
\newblock \doi{10.1017/cbo9780511551512}
\bibAnnoteFile{Gary1993}

\bibitem[{Gary and Tokar(1985)}]{Gary1985}
Gary, S.~P. and Tokar, R.~L. (1985).
\newblock The electron-acoustic mode.
\newblock \emph{Phys Fluids} 28, 2439.
\newblock \doi{10.1063/1.865250}
\bibAnnoteFile{Gary1985}

\bibitem[{Guo(2020)}]{Guo2020}
Guo, R. (2020).
\newblock Stationary states of polytropic plasmas.
\newblock \emph{Phys Plasmas} 27, 122104.
\newblock \doi{10.1063/5.0024222}
\bibAnnoteFile{Guo2020}

\bibitem[{Guo(2021{\natexlab{a}})}]{Guo2021a}
Guo, R. (2021{\natexlab{a}}).
\newblock The electron acoustic waves in plasmas with two kappa-distributed
  electrons at the same temperatures and immobile ions.
\newblock \emph{Phys Plasmas} 28, 082105.
\newblock \doi{10.1063/5.0057693}
\bibAnnoteFile{Guo2021a}

\bibitem[{Guo(2021{\natexlab{b}})}]{Guo2021}
Guo, R. (2021{\natexlab{b}}).
\newblock An improved approach to derive the kappa distribution in polytropic
  plasmas.
\newblock \emph{Phys Plasmas} 28, 044502.
\newblock \doi{10.1063/5.0041522}
\bibAnnoteFile{Guo2021}

\bibitem[{Guo and Du(2019)}]{Guo2019}
Guo, R. and Du, J. (2019).
\newblock Transport coefficients of the fully ionized plasma with
  kappa-distribution and in strong magnetic field.
\newblock \emph{Physica A} 523, 156--171.
\newblock \doi{10.1016/j.physa.2019.02.011}
\bibAnnoteFile{Guo2019}

\bibitem[{Hapgood et~al.(2011)Hapgood, Perry, Davies, and Denton}]{Hapgood2011}
Hapgood, M., Perry, C., Davies, J., and Denton, M. (2011).
\newblock The role of suprathermal particle measurements in {CrossScale}
  studies of collisionless plasma processes.
\newblock \emph{Planet Space Sci} 59, 618--629.
\newblock \doi{10.1016/j.pss.2010.06.002}
\bibAnnoteFile{Hapgood2011}

\bibitem[{Hatami et~al.(2018)Hatami, Tribeche, and Mamun}]{Hatami2018}
Hatami, M.~M., Tribeche, M., and Mamun, A.~A. (2018).
\newblock Debye length and electric potential in magnetized nonextensive
  plasma.
\newblock \emph{Phys Plasmas} 25, 094502.
\newblock \doi{10.1063/1.5036760}
\bibAnnoteFile{Hatami2018}

\bibitem[{Hellberg et~al.(2009)Hellberg, Mace, Baluku, Kourakis, and
  Saini}]{Hellberg2009}
Hellberg, M.~A., Mace, R.~L., Baluku, T.~K., Kourakis, I., and Saini, N.~S.
  (2009).
\newblock Comment on mathematical and physical aspects of kappa velocity
  distribution [phys. plasmas 14, 110702 (2007)].
\newblock \emph{Phys Plasmas} 16, 094701.
\newblock \doi{10.1063/1.3213388}
\bibAnnoteFile{Hellberg2009}

\bibitem[{Husidic et~al.(2021)Husidic, Lazar, Fichtner, Scherer, and
  Poedts}]{Husidic2021}
Husidic, E., Lazar, M., Fichtner, H., Scherer, K., and Poedts, S. (2021).
\newblock Transport coefficients enhanced by suprathermal particles in
  nonequilibrium heliospheric plasmas.
\newblock \emph{Astronomy {\&} Astrophysics} 654, A99.
\newblock \doi{10.1051/0004-6361/202141760}
\bibAnnoteFile{Husidic2021}

\bibitem[{Kasaba et~al.(2000)Kasaba, Terasawa, Tsubouchi, Mukai, Saito,
  Matsumoto et~al.}]{Kasaba2000}
Kasaba, Y., Terasawa, T., Tsubouchi, K., Mukai, T., Saito, Y., Matsumoto, H.,
  et~al. (2000).
\newblock Magnetosheath electrons in anomalously low density solar wind
  observed by geotail.
\newblock \emph{Geophys Res Lett} 27, 3253--3256.
\newblock \doi{10.1029/2000gl000086}
\bibAnnoteFile{Kasaba2000}

\bibitem[{Lazar et~al.(2016)Lazar, Fichtner, and Yoon}]{Lazar2016}
Lazar, M., Fichtner, H., and Yoon, P.~H. (2016).
\newblock On the interpretation and applicability of $\kappa$-distributions.
\newblock \emph{Astronomy {\&} Astrophysics} 589, A39.
\newblock \doi{10.1051/0004-6361/201527593}
\bibAnnoteFile{Lazar2016}

\bibitem[{Lazar et~al.(2018)Lazar, Kourakis, Poedts, and Fichtner}]{Lazar2018}
Lazar, M., Kourakis, I., Poedts, S., and Fichtner, H. (2018).
\newblock On the effects of suprathermal populations in dusty plasmas: The case
  of dust-ion-acoustic waves.
\newblock \emph{Planet Space Sci} 156, 130--138.
\newblock \doi{10.1016/j.pss.2017.11.011}.
\newblock Dust, Atmosphere, and Plasma Environment of the Moon and Small Bodies
\bibAnnoteFile{Lazar2018}

\bibitem[{Lazar et~al.(2017)Lazar, Pierrard, Shaaban, Fichtner, and
  Poedts}]{Lazar2017a}
Lazar, M., Pierrard, V., Shaaban, S.~M., Fichtner, H., and Poedts, S. (2017).
\newblock Dual maxwellian-kappa modeling of the solar wind electrons: new clues
  on the temperature of kappa populations.
\newblock \emph{Astronomy {\&} Astrophysics} 602, A44.
\newblock \doi{10.1051/0004-6361/201630194}
\bibAnnoteFile{Lazar2017a}

\bibitem[{Livadiotis(2015)}]{Livadiotis2015a}
Livadiotis, G. (2015).
\newblock Introduction to special section on origins and properties of kappa
  distributions: Statistical background and properties of kappa distributions
  in space plasmas.
\newblock \emph{J. Geophys. Res. Space Phys.} 120, 1607--1619.
\newblock \doi{10.1002/2014ja020825}
\bibAnnoteFile{Livadiotis2015a}

\bibitem[{Livadiotis(2019{\natexlab{a}})}]{Livadiotis2019}
Livadiotis, G. (2019{\natexlab{a}}).
\newblock On the generalized formulation of debye shielding in plasmas.
\newblock \emph{Phys Plasmas} 26, 050701.
\newblock \doi{10.1063/1.5091949}
\bibAnnoteFile{Livadiotis2019}

\bibitem[{Livadiotis(2019{\natexlab{b}})}]{Livadiotis2019b}
Livadiotis, G. (2019{\natexlab{b}}).
\newblock On the origin of polytropic behavior in space and astrophysical
  plasmas.
\newblock \emph{Astrophys. J.} 874, 10.
\newblock \doi{10.3847/1538-4357/ab05b7}
\bibAnnoteFile{Livadiotis2019b}

\bibitem[{Livadiotis and McComas(2010)}]{Livadiotis2010a}
Livadiotis, G. and McComas, D.~J. (2010).
\newblock Exploring transitions of space plasmas out of equilibrium.
\newblock \emph{Astrophys. J.} 714, 971--987.
\newblock \doi{10.1088/0004-637X/714/1/971}
\bibAnnoteFile{Livadiotis2010a}

\bibitem[{Mace et~al.(1999)Mace, Amery, and Hellberg}]{Mace1999}
Mace, R.~L., Amery, G., and Hellberg, M.~A. (1999).
\newblock The electron-acoustic mode in a plasma with hot suprathermal and cool
  maxwellian electrons.
\newblock \emph{Phys Plasmas} 6, 44--49.
\newblock \doi{10.1063/1.873256}
\bibAnnoteFile{Mace1999}

\bibitem[{Mace and Hellberg(1995)}]{Mace1995}
Mace, R.~L. and Hellberg, M.~A. (1995).
\newblock A dispersion function for plasmas containing superthermal particles.
\newblock \emph{Phys Plasmas} 2, 2098--2109.
\newblock \doi{10.1063/1.871296}
\bibAnnoteFile{Mace1995}

\bibitem[{Myllys et~al.(2019)Myllys, Henri, Galand, Heritier, Gilet, Goldstein
  et~al.}]{Myllys2019}
Myllys, M., Henri, P., Galand, M., Heritier, K.~L., Gilet, N., Goldstein, R.,
  et~al. (2019).
\newblock Plasma properties of suprathermal electrons near comet
  67p/churyumov-gerasimenko with rosetta.
\newblock \emph{Astronomy {\&} Astrophysics} 630, A42.
\newblock \doi{10.1051/0004-6361/201834964}
\bibAnnoteFile{Myllys2019}

\bibitem[{Ogasawara et~al.(2017)Ogasawara, Livadiotis, Grubbs, Jahn, Michell,
  Samara et~al.}]{Ogasawara2017}
Ogasawara, K., Livadiotis, G., Grubbs, G.~A., Jahn, J.-M., Michell, R., Samara,
  M., et~al. (2017).
\newblock Properties of suprathermal electrons associated with discrete auroral
  arcs.
\newblock \emph{Geophys Res Lett} 44, 3475--3484.
\newblock \doi{10.1002/2017gl072715}
\bibAnnoteFile{Ogasawara2017}

\bibitem[{Pierrard et~al.(2016)Pierrard, Lazar, Poedts, {\v{S}}tver{\'{a}}k,
  Maksimovic, and Tr{\'{a}}vn{\'{\i}}{\v{c}}ek}]{Pierrard2016}
Pierrard, V., Lazar, M., Poedts, S., {\v{S}}tver{\'{a}}k, {\v{S}}., Maksimovic,
  M., and Tr{\'{a}}vn{\'{\i}}{\v{c}}ek, P.~M. (2016).
\newblock The electron temperature and anisotropy in the solar wind. comparison
  of the core and halo populations.
\newblock \emph{Sol Phys} 291, 2165--2179.
\newblock \doi{10.1007/s11207-016-0961-7}
\bibAnnoteFile{Pierrard2016}

\bibitem[{Pottelette et~al.(1999)Pottelette, Ergun, Treumann, Berthomier,
  Carlson, McFadden et~al.}]{Pottelette1999}
Pottelette, R., Ergun, R.~E., Treumann, R.~A., Berthomier, M., Carlson, C.~W.,
  McFadden, J.~P., et~al. (1999).
\newblock Modulated electron-acoustic waves in auroral density cavities: {FAST}
  observations.
\newblock \emph{Geophys Res Lett} 26, 2629--2632.
\newblock \doi{10.1029/1999gl900462}
\bibAnnoteFile{Pottelette1999}

\bibitem[{Singh et~al.(2001)Singh, Reddy, and Lakhina}]{Singh2001a}
Singh, S., Reddy, R., and Lakhina, G. (2001).
\newblock Broadband electrostatic noise due to nonlinear electron-acoustic
  waves.
\newblock \emph{Adv Space Res} 28, 1643--1648.
\newblock \doi{10.1016/s0273-1177(01)00479-3}
\bibAnnoteFile{Singh2001a}

\bibitem[{Summers and Thorne(1991)}]{Summers1991}
Summers, D. and Thorne, R.~M. (1991).
\newblock The modified plasma dispersion function.
\newblock \emph{Phys. Fluids B} 3, 1835--1847.
\newblock \doi{10.1063/1.859653}
\bibAnnoteFile{Summers1991}

\bibitem[{Wang et~al.(2021)Wang, Du, and Huo}]{Wang2021}
Wang, H., Du, J., and Huo, R. (2021).
\newblock The collision frequency of electron-neutral-particle in weakly
  ionized plasmas with non-maxwellian velocity distributions.
\newblock \emph{Commun Theor Phys} 73, 095501.
\newblock \doi{10.1088/1572-9494/ac0a6f}
\bibAnnoteFile{Wang2021}

\bibitem[{Wang and Du(2018)}]{Wang2018}
Wang, Y. and Du, J. (2018).
\newblock The viscosity of charged particles in the weakly ionized plasma with
  power-law distributions.
\newblock \emph{Phys Plasmas} 25, 062309.
\newblock \doi{10.1063/1.5023030}
\bibAnnoteFile{Wang2018}

\bibitem[{Yoon(2014)}]{Yoon2014}
Yoon, P.~H. (2014).
\newblock Electron kappa distribution and quasi-thermal noise.
\newblock \emph{J. Geophys. Res. Space Phys.} 119, 7074--7087.
\newblock \doi{10.1002/2014ja020353}
\bibAnnoteFile{Yoon2014}

\end{thebibliography}
\end{document}